\documentclass[prd,aps,showpacs,nofootinbib,preprint,eqsecnum]{revtex4}
\usepackage{graphicx}
\usepackage{color}
\usepackage{amssymb,amsmath}

\usepackage{amsxtra}
\usepackage{epsf}
\usepackage{amssymb}
\usepackage{enumerate}
\usepackage{hhline}
\usepackage{array}
\usepackage{tabularx}
\newcommand{\e}{\mathrm{e}}
\newcommand{\Eqn}[1]{&\hspace{-0.2em}#1\hspace{-0.2em}&}

\begin{document}

\title{Reconstruction of the equation of state for the cyclic universes in homogeneous and isotropic cosmology} 


\author{
Kazuharu Bamba$^{1, 2,}$\footnote{
E-mail address: bamba@kmi.nagoya-u.ac.jp}, 
Kuralay Yesmakhanova$^{2}$, 
Koblandy Yerzhanov$^{2}$ 
and 
Ratbay Myrzakulov$^{2,}$\footnote{
E-mail addresses: rmyrzakulov@gmail.com}}
\affiliation{
$^1$Kobayashi-Maskawa Institute for the Origin of Particles and the
Universe,
Nagoya University, Nagoya 464-8602, Japan\\
$^2$\textit{Eurasian International Center for Theoretical Physics,} \\ \textit{Eurasian National University, Astana 010008, Kazakhstan}}

%
%


\begin{abstract} 
We study the cosmological evolutions of the equation of state (EoS) for the 
universe in the homogeneous and isotropic Friedmann-Lema\^{i}tre-Robertson-Walker (FLRW) space-time. 
In particular, we reconstruct the cyclic universes by using the Weierstrass 
and Jacobian elliptic functions. 
It is explicitly illustrated that in several models the universe always stays 
in the non-phantom (quintessence) phase, whereas there also exist models 
in which the crossing of the phantom divide can be realized in the reconstructed cyclic universes. 
\end{abstract}



\maketitle

\section{Introduction} 

Recent observations on cosmic microwave background (CMB) radiation~\cite{WMAP} 
have suggested that in the early universe inflation occurred and the universe became homogeneous and isotropic. 
On the other hand, according to various recent cosmological observations such as Type Ia Supernovae~\cite{SN1}, CMB radiation~\cite{WMAP}, 
large scale structure (LSS)~\cite{LSS}, 
baryon acoustic oscillations (BAO)~\cite{Eisenstein:2005su}, 
and weak lensing~\cite{Jain:2003tba}, 
the current cosmic expansion is accelerating. 
Hence, in the present time so-called dark energy, 
which may be some unknown matter or geometrical energy coming from 
the deviation of a gravitational theory from general relativity, 
is dominant over non-relativistic matter, i.e., cold dark matter, 
radiation, and baryon (for reviews on dark energy, see, e.g.,~\cite{
Copeland:2006wr, D-M, Cai:2009zp, Tsujikawa:2010sc, 
Book-Amendola-Tsujikawa, Li:2011sd, Bamba:2012cp}; for those on modifications of gravity, 
see, e.g.,~\cite{Review-Nojiri-Odintsov, 
Sotiriou:2008rp, Book-Capozziello-Faraoni, Capozziello:2011et, 
DeFelice:2010aj, Clifton:2011jh, Capozziello:2012hm}). 

In the beginning of the universe before inflation, it is considered that 
there was so-called a Big Bang singularity. In other words, the universe might 
have evolved from a singular space-time point. 
On the other hand, it is known that at dark energy dominant stage, 
there eventually appear the finite-time future singularities~\cite{Big-Rip, Shtanov:2002ek, Barrow:2004xh, sudden, Nojiri:2005sx, Future-singularity-MG}, 
or there are several scenarios in which the late time universe contracts and eventually a Big Crunch singularity occurs. 
Thus, the issue discussed here is that there exist singularities in the 
beginning of the universe as well as at the last stage of it. 
If the evolution of the universe is periodic, the existence of 
a Big Bang singularity as well as the finite-time future singularities or a Big Crunch can be avoided. Based on this idea and inspired by string theories, the cyclic universe has been proposed in Ref.~\cite{Cyclic-universe}\footnote{In other context, the cyclic universe has been argued in Ref.~\cite{Chung:2001ka}} (for recent comprehensive papers on the cyclic universe, see, e.g.,~\cite{Recent-Cyclic-Papers, Cai:2011bs, BMKG-C-U, Sahni:2012er}). In addition, the ekpyrotic scenario in the framework of the brane world has also been suggested in Ref.~\cite{Ekpyrotic-scenarios}. 
Furthermore, the bouncing universe has been investigated in Refs.~\cite{Bouncing-cosmology, Corda:2010ni} (for a review on bouncing cosmology, see~\cite{Novello:2008ra}). 
Moreover, in Ref.~\cite{Knot-universe} the (trefoil and figure-eight) knot universe has been studied, where the knot theory relates to the cyclic universe. In other words, the geometrical picture of the knot relates to oscillatory solutions of the gravitational field equations in the homogeneous and isotropic Friedmann-Lema\^{i}tre-Robertson-Walker (FLRW) and Bianchi-type I universes. 
Recently, there has also executed an investigation of figure eight knot in 
Ref.~\cite{Itoyama:2012fq}. 
Furthermore, it is suggested~\cite{Elliptic-functions, K4} that the Weierstrass $\wp(t)$, $\zeta(t)$ and $\sigma(t)$-functions and the Jacobian elliptic functions can play an important role to examine astrophysical and cosmological 
problems (for recent studies of the applications, see, e.g.,~\cite{R-E-S}). In particular, in Ref.~\cite{K4} the elliptic functions have been applied to describe the FLRW universe. 
Related cosmological features such as a cyclic behavior of the cosmic 
evolution have been studied in various scenarios~\cite{MGA}.
Moreover, in Ref.~\cite{Bamba:2012yy}, 
the behavior of the EoS for dark energy 
has been investigated in the so-called g-essence models 
constructed by both k-essence~\cite{k-essence} and f-essence~\cite{Myrzakulov:2010du}, 
which is treated as a spinor field and 
corresponds to a classical c-number quantity. 
More recently, 
f-essence is dealt with a Grassmann-valued
quantity in Ref.~\cite{Damour:2009zc} or an operator, i.e., q-number 
in Ref.~\cite{Damour:2011yk}. 
Generalization of the Chaplygin gas type models~\cite{C-G-T-M} with the periodicity or the quasi-periodicity have also been explored in Ref.~\cite{Bamba:2012wb}.

In this paper, with the Weierstrass and Jacobian elliptic functions, we reconstruct the cyclic universe 
by using the equivalent procedure in Refs.~\cite{Review-Nojiri-Odintsov, Nojiri:2005sx, Nojiri:2005sr, Stefancic:2004kb}. 
It is important to emphasize that 
to use the Weierstrass and Jacobian elliptic functions for describing the 
scale factor or the Hubble parameter 
is one of the novel ingredients in this work. 
The cosmological motivation to use such elliptic functions for the 
scale factor or the Hubble parameter is to realize a cyclic behavior of the 
universe naturally without special setting 
by using the properties of periodicities of the Weierstrass and Jacobian elliptic functions. This corresponds to the reconstruction procedures, through which 
an arbitrary cosmological expansion history can be reconstructed by 
starting with providing an appropriate form of the 
scale factor or the Hubble parameter with desirable features. 
We use the units of 
the gravitational constant 
$8 \pi G = c =1$ with $G$ and $c$ being the gravitational constant and the seed of light. 

The paper is organized as follows. 
In Sec.\ II, we explain the basic equations in the FLRW space-time. 
In Sec.\ III, we explore models induced by the Weierstrass $\zeta(t)$-function.  
In Sec.\ IV, we examine models induced by the Weierstrass $\sigma(t)$-function.  
In Sec.\ V, we investigate models induced by Jacobian elliptic functions. 
Finally, conclusions are given in Sec.\ VI. 

\section{FLRW cosmology} 

We start with the standard gravitational action
\begin{equation}
S=\int\sqrt{-g}d^4x \left( R+L_{\mathrm{m}} \right)\,, 
\label{eq:2.1}
\end{equation}
where
$R$ is the scalar curvature and $L_{\mathrm{m}}$ is the Lagrangian of matter. 
Now, we assume the FLRW space-time with the metric, 
\begin{equation}
ds^2=-dt^2+a^2(t)\left[\frac{dr^2}{1-Kr^2}+r^2d\Omega^2\right]\,,
\label{eq:2.2}
\end{equation}
where $a(t)$ is the scale factor, 
$d \Omega^2$ is the metric of 2-dimensional sphere with unit radius. 
Moreover, $K$ is the cosmic curvature, and 
for $K=-1$, $K=0$, and $K=+1$\footnote{$K$ can take any value, but it is related to $(-, 0, +)$ curvatures according to its sign.}, the universe is open, flat, and closed, respectively. 
Moreover, the Hubble parameter 
is defined by $H \equiv \dot{a}/a$, where a dot denotes the time derivative, 
$\partial/\partial t$. 
In the flat FLRW background (\ref{eq:2.2}), that is, $k=0$, from the action 
in Eq.~(\ref{eq:2.1}), the gravitational equations and the continuity equation can be written in 
\textit{the H-form} 
\begin{eqnarray}
\rho\Eqn{=}3H^2\,,
\label{eq:2.4} \\
P\Eqn{=}-2\dot{H}-3H^2\,,
\label{eq:2.3} \\
\dot{\rho}\Eqn{=}-3H \left( \rho+P \right)\,, 
\end{eqnarray}
where $\rho$ and $P$ are the energy density and pressure, respectively.

\section{Models induced by the Weierstrass $\zeta(t)$-function}

In this section, we study a model (MG-II) 
induced by the Weierstrass $\zeta(t)$-function. 
In particular, the Weierstrass $\zeta(t)$-function is the logarithmic 
derivative of the $\sigma(t)$-function, which we use in Sec.~IV. 
We use the following procedure for the reconstruction of the EoS. 
First, we express the scale factor or the Hubble parameter by 
using the Weierstrass $\zeta(t)$-function. 
Next, by combining these expressions with the gravitational field equations (\ref{eq:2.3}) and (\ref{eq:2.4}), we derive the energy density and pressure, 
and thus reconstruct the EoS. 
In this paper, the so-called Myrzakulov Gas (MG)-$i$ ($i$ = I, II, $\dots$, XV) model means the model of gases or fluid, 
according to the notations of Ref.~\cite{Knot-universe}. 
We note that 
the Weierstrass $\zeta(t)$-function is a quasi-periodic 
function and 
the Weierstrass $\wp(t)$-function is a two periodic function. 
Thus, 
in comparison with the use of ordinary trigonometric functions, 
the advantage of the use of 
the Weierstrass $\wp(t)$-function
is that two periodic behaviors can be obtained. 

The physical motivation why 
we examine the MG-$i$ gas 
is that in these models the cosmological evolution 
of the generalized Chaplygin gas type models with 
the periodical and quasi-periodical features can be realized. 
The detailed behaviors are dependent on the models. 
The important point is that 
these models described by the Weierstrass functions. 
Therefore, the cosmic expansion history showing 
the periodicity and/or quasi-periodicity behaviors 
can be demonstrated. 
As a consequence, 
these models can produce new cosmological scenarios in which 
cosmological singularities such as 
a Big Bang singularity, the finite-time future singularities or 
a Big Crunch singularity can be removed, 
similarly to that in, e.g., 
the cyclic universe, ekpyrotic and bouncing universe scenarios. 

\subsection{MG-II model}

We investigate the MG-II model when 
the scale factor is written 
with the $\zeta(t)$-function as follows 
\begin{equation}
a=\zeta (t)\,, 
\label{eq:3.6}
\end{equation}
where $\zeta (t)$ is the $\zeta$-Weierstrass function. 
Hence, 
the Hubble parameter takes the form 
\begin{equation}
H=-\frac{\wp(t)}{\zeta (t)}\,.
\end{equation}
where $\wp(t)$ is the $\wp$-Weierstrass function. 
In this case, the parametric EoS is given by 
\begin{eqnarray}
\rho \Eqn{=} 3(\frac{\wp(t)}{\zeta(t)})^2\,,
\label{eq:3.9-1} \\
P\Eqn{=}-\frac{\wp^2(t)-2\zeta(t)\acute{\wp}(t)}{\zeta^2(t)}\,,
\label{eq:3.8-1}
\end{eqnarray}
where a prime denotes the time derivative of $d/dt$. 
The EoS parameter is expressed by
\begin{equation}
w \equiv \frac{P}{\rho} = -1 + \frac{2}{3} \left( 1 +\frac{\acute{\wp}(t)}{\zeta(t)} \right)\,.
\label{eq:3.10-1}
\end{equation}
%

\begin{center}
\begin{figure}[t]
\resizebox{!}{6.5cm}{
   \includegraphics{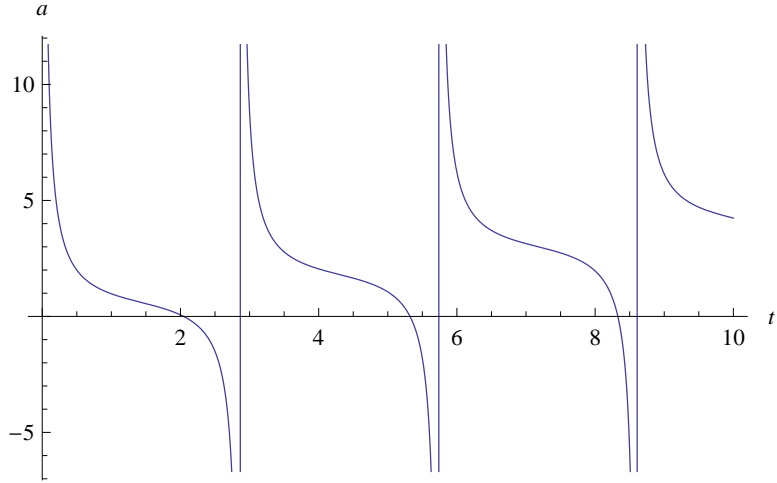}
                  }
\caption{The scale factor $a$ in Eq.~(\ref{eq:3.6}) 
as a function of $t$ in the MG-II model for 
the Weierstrass invariants of $g_1 =1$ and $g_2 =1$.}
\label{fig-1}
\end{figure}
\end{center}

\begin{center}
\begin{figure}[t]
\begin{tabular}{ll}
\begin{minipage}{80mm}
\begin{center}
\unitlength=1mm
\resizebox{!}{4.5cm}{
   \includegraphics{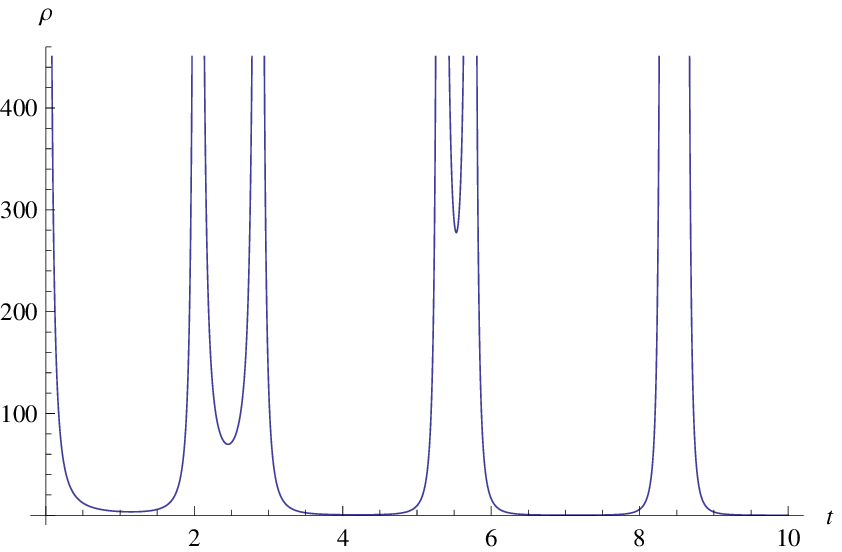}
                  }
\end{center}
\end{minipage}
&
\begin{minipage}{80mm}
\begin{center}
\unitlength=1mm
\resizebox{!}{4.5cm}{
   \includegraphics{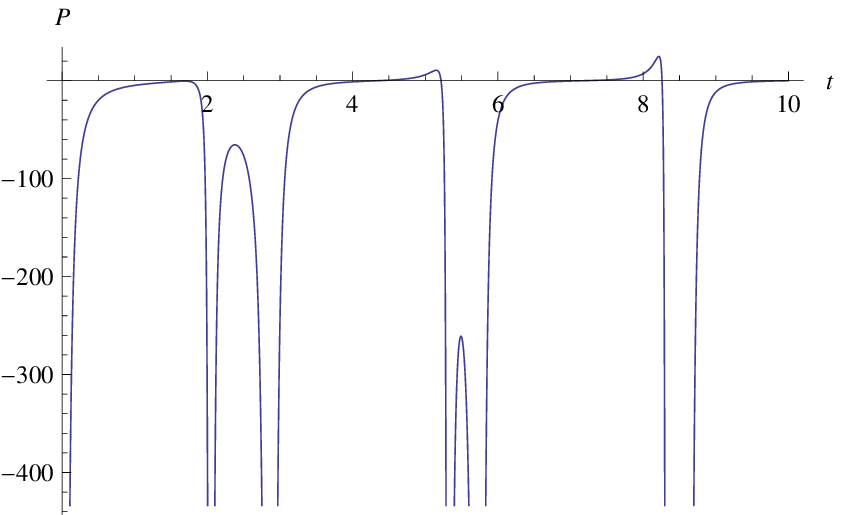}
                  }
\end{center}
\end{minipage}

\end{tabular}
\caption{The energy density $\rho$ in Eq.~(\ref{eq:3.9-1}) [left panel] 
and pressure $P$ in Eq.~(\ref{eq:3.8-1}) [right panel] 
as functions of $t$. Legend is the same as Fig.~\ref{fig-1}.
}
\label{fig-2}
\end{figure}
\end{center}

\begin{center}
\begin{figure}[t]
\resizebox{!}{6.5cm}{
   \includegraphics{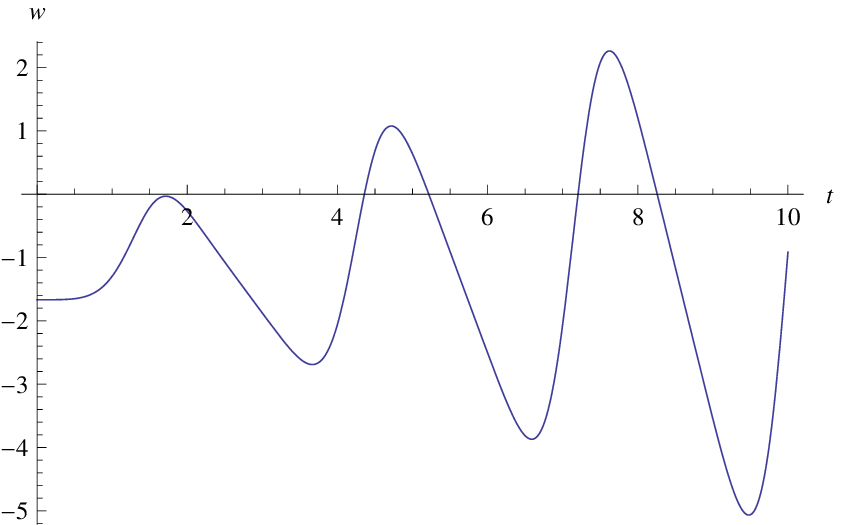}
                  }
\caption{The EoS $w$ in Eq.~(\ref{eq:3.10-1}) as a function of $t$. Legend is the same as Fig.~\ref{fig-1}.}
\label{fig-3}
\end{figure}
\end{center}

In Fig.~\ref{fig-1}, 
we show the cosmological evolution of the scale factor $a(t)$ 
in Eq.~(\ref{eq:3.6}) as a function of $t$. We also depict the cosmological evolutions of the energy density $\rho$ in Eq.~(\ref{eq:3.9-1}) 
and pressure $P$ in 
Eq.~(\ref{eq:3.8-1}) as functions of $t$ in Fig.~\ref{fig-2}. 
Furthermore, in Fig.~\ref{fig-3} we demonstrate the cosmological evolution of 
the EoS $w$ in Eq.~(\ref{eq:3.10-1}) as a function of $t$. 
Here, we have used the Weierstrass invariants of $g_1 =1$ and $g_2 =1$, 
which are defined to satisfy the following equation~\cite{K4} 
\begin{equation}
\wp^{-1}(t; g_1, g_2) = \int_{t}^{\infty} \frac{1}{\sqrt{4 \left(t^{\prime}\right)^3 - g_1 t^{\prime} - g_2}}d t^{\prime}\,.
\label{eq:Bamba-Add-3-01}
\end{equation} 
{}From Fig.~\ref{fig-1}, we see the oscillatory behavior of $a$. 
Thus, it is interpreted that by using the Weierstrass 
$\zeta(t)$-function, a model describes the cyclic universe with two periods. 
This comes from the feature of the two periodicity of the Weierstrass 
$\zeta(t)$-function. 
We note that in Figs.~\ref{fig-1} and \ref{fig-2}, there are no diverging 
behavior of $a$, $\rho$ and $p$, namely, all the curves in each figure are 
smoothly connected\footnote{Since the amplitude is very large, in Figs.~\ref{fig-1} and \ref{fig-2} apparently it seems there are some divergence. Indeed, however, there are no divergence.}. 

On the other hand, the effective EoS for the universe 
in the FLRW space-time (2.2) is described by~\cite{Review-Nojiri-Odintsov} 
$
w_{\mathrm{eff}} \equiv P_{\mathrm{eff}}/\rho_{\mathrm{eff}} = 
-1 - 2\dot{H}/\left(3H^2\right)
$ with  
$
\rho_{\mathrm{eff}} \equiv 3H^2
$ and 
$
P_{\mathrm{eff}} \equiv -\left(2\dot{H}+3H^2\right)
$,  
where 
$\rho_{\mathrm{eff}}$ and $P_{\mathrm{eff}}$ correspond to 
the total energy density and pressure of the universe, respectively. 
Throughout this paper, $\rho_{\mathrm{eff}}$ is $\rho$ in Eq.~(2.2) 
and $p_{\mathrm{eff}}$ is $P$ in Eq.~(2.3). 
For $\dot{H} < 0$, $w_\mathrm{eff} >-1$, which is the non-phantom (quintessence) phase, whereas for $\dot{H} > 0$, $w_\mathrm{eff} <-1$, 
which is the phantom phase. 
For $\dot{H} = 0$, $w_\mathrm{eff} =-1$, which corresponds to the 
cosmological constant. 
It is significant to remark that 
recent various observational data~\cite{observational status} implies that 
the crossing of the phantom divide line of $w_{\mathrm{DE}}=-1$ occurred 
in the near past. Here, $w_{\mathrm{DE}}$ is the EoS for dark energy 
and at the dark energy dominated stage, it can be regarded that 
$w \approx w_{\mathrm{DE}} \approx w_{\mathrm{eff}}$. 
{}From Fig.~\ref{fig-3}, we find that multiple crossings of the phantom divide 
can be realized. 

We note that if the two periods of the elliptic functions are equal to 
infinity, that is, in the case $m_1=m_2=\infty$, the elliptic functions are 
reduced (degenerate) to the elementary rational functions: 
\begin{equation}
\sigma(t)=t\,, 
\quad 
\zeta(t)=t^{-1}\,, 
\quad \wp(t)=t^{-2}\,. 
\label{eq:3.7}
\end{equation}
In this degenerate case, the MG-II model is reduced to the case 
\begin{equation}
a=-H=\frac{1}{t} 
\label{eq:3.8-2}
\end{equation}
At the same time, the parametric EoS takes the form 
\begin{equation}
\rho = \frac{3}{t^2}\,, 
\quad 
P=-\frac{5}{t^2}\,. 
\label{eq:3.9-2}
\end{equation}
This means that the EoS reads 
\begin{equation}
P=-\frac{5\rho}{3}\,.
\label{eq:3.10-2}
\end{equation}
Hence, the corresponding EoS parameter is given by
\begin{equation}
w =-\frac{5}{3}\approx -1.67\,,
\label{eq:3.11}
\end{equation}
so that the crossing of the phantom divide with $w \approx -1.67$ can 
be realized. 
Thus, we can conclude that the MG-II model is the two-periodical analogue of the usual universe filled by the barotropic fluid with the EoS of the form (\ref{eq:3.10-2}) and that the crossing of the phantom divide with (\ref{eq:3.11}) 
occurs. 

\section{Models induced by the Weierstrass $\sigma(t)$-function}

In this section, we explore two models (MG-I, MG-III) 
induced by the Weierstrass $\sigma(t)$-function, 
which is a quasi-periodic function. 
We reconstruct the EoS by using the same procedure as that in Sec.~III.

\subsection{MG-I model}

We represent the scale factor as 
\begin{equation}
a=a_0\sigma(t)\,,
\label{eq:4.1}
\end{equation}
where $a_0$ is a constant and 
$\sigma(t)$ is the $\sigma$-Weierstrass function. 
In this case, the Hubble parameter is described by 
\begin{equation}
H(t)=\zeta(t)\,. 
\end{equation}
Then, the gravitational field equations lead to 
the parametric EoS as 
\begin{eqnarray}
\rho \Eqn{=} 3\zeta^2(t)\,,
\label{eq:4.4} \\
P \Eqn{=} 2\wp(t)-3\zeta^2(t)\,,
\label{eq:4.3} 
\end{eqnarray} 
and the EoS parameter is given by 
\begin{equation}
w = -1+\frac{2}{3}\frac{\wp(t)}{\zeta^2(t)}\,.
\label{eq:4.5}
\end{equation}
%

\begin{center}
\begin{figure}[t]
\resizebox{!}{6.5cm}{
   \includegraphics{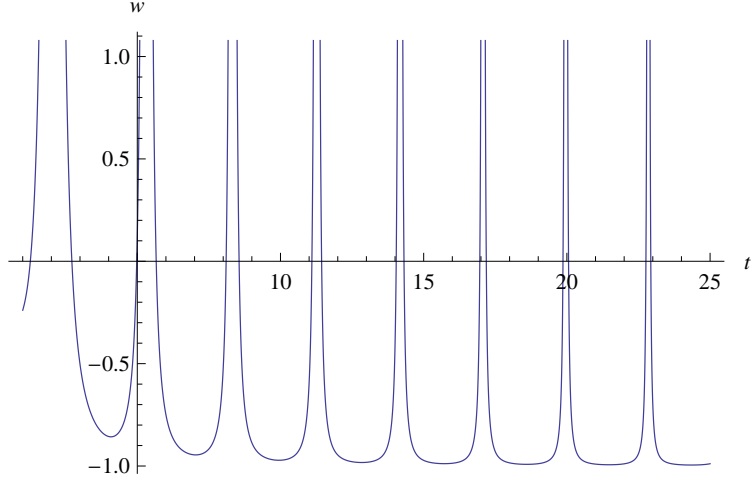}
                  }
\caption{The EoS $w$ in Eq.~(\ref{eq:4.5}) as a function of $t$. Legend is the same as Fig.~\ref{fig-1}.}
\label{fig-4}
\end{figure}
\end{center}

In Fig.~\ref{fig-4}, we display the cosmological evolution of 
the EoS $w$ in Eq.~(\ref{eq:4.5}) as a function of $t$. 
Here, we have used the Weierstrass invariants of $g_1 =1$ and $g_2 =1$ 
in Eq.~(\ref{eq:Bamba-Add-3-01}). 
It follows from Fig.~\ref{fig-4} that the universe always stays 
in the non-phantom phase ($w > -1$). 

We also examine the case that two periods of the elliptic functions are equal 
to infinity, that is, in the case $m_1=m_2=\infty$, the elliptic functions are 
reduced (degenerate) to the elementary rational functions according to the equations in \eqref{eq:3.7}. 
In this degenerate case, the MG-I model is reduced to the case in which 
the scale factor and the Hubble parameter are given by
\begin{equation}
a=a_0t\,, 
\quad 
H=\frac{1}{t}\,. 
\label{eq:4.6}
\end{equation}
At the same time, the parametric EoS is expressed as 
\begin{equation}
\rho = \frac{3}{t^2}\,, 
\quad 
P=-\frac{1}{t^2}\,. 
\label{eq:4.7}
\end{equation}
Therefore, the EoS is described as 
\begin{equation}
P=-\frac{\rho}{3}\,.
\label{eq:4.8}
\end{equation}
Hence, the corresponding EoS parameter becomes 
\begin{equation}
w =-\frac{1}{3}\,. 
\label{eq:4.9}
\end{equation}
As a result, it can be concluded 
that the MG-I model is the two-periodic analogue of the usual universe filled with the barotropic fluid with the EoS in Eq.~(\ref{eq:4.8}) and 
with the EoS parameter in Eq.~(\ref{eq:4.9}). 

\subsection{MG-III model}

\begin{center}
\begin{figure}[t]
\begin{tabular}{ll}
\begin{minipage}{80mm}
\begin{center}
\unitlength=1mm
\resizebox{!}{4.5cm}{
   \includegraphics{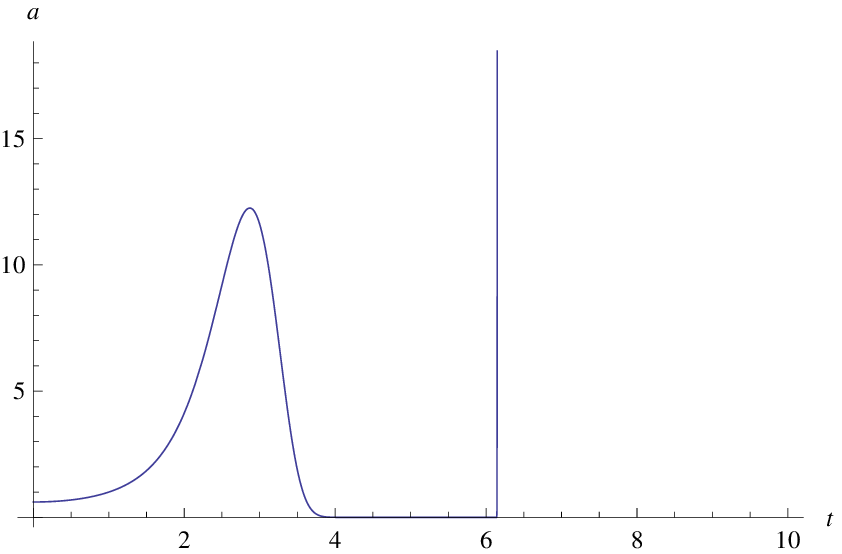}
                  }
\end{center}
\end{minipage}
&
\begin{minipage}{80mm}
\begin{center}
\unitlength=1mm
\resizebox{!}{4.5cm}{
   \includegraphics{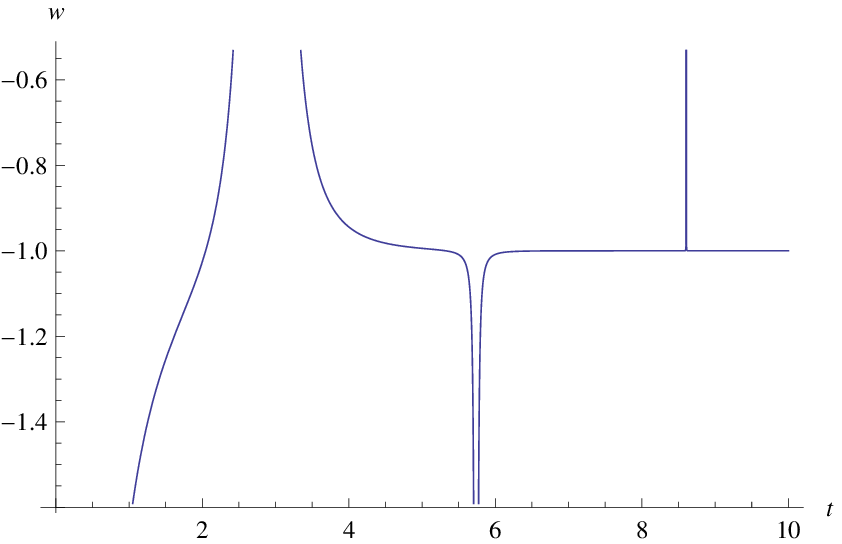}
                  }
\end{center}
\end{minipage}

\end{tabular}
\caption{The scale factor $a$ in Eq.~(\ref{eq:III-a}) [left panel] and the EoS $w$ in Eq.~(\ref{eq:III-w}) [right panel] as functions of $t$ in the MG-III model. Legend is the same as Fig.~\ref{fig-1}.
}
\label{fig-5}
\end{figure}
\end{center}

Next, we express the Hubble parameter as 
\begin{equation}
H(t)=\sigma(t)\,.
\end{equation}
{}From this representation of $H$, the scale factor is given by 
\begin{equation}
a=a_0\exp \left( \int^{t}_{1}\sigma(K)dK \right)\,. 
\label{eq:III-a}
\end{equation}
Thus, the gravitational field equations give 
the parametric EoS as 
\begin{eqnarray}
\rho \Eqn{=} 3\sigma^2(t)\,, \\ 
P \Eqn{=} -2\sigma(t)\zeta(t)-3\sigma^2(t)\,,
\end{eqnarray}
Moreover, 
the EoS parameter becomes 
\begin{equation}
w = -1-\frac{2}{3}\frac{\zeta(t)}{\sigma(t)}\,.
\label{eq:III-w}
\end{equation}
In Fig.~\ref{fig-5}, 
we show the cosmological evolutions of 
the scale factor $a$ in Eq.~(\ref{eq:III-a}) and the EoS $w$ in 
Eq.~(\ref{eq:III-w}) as functions of $t$ for the Weierstrass invariants of 
$g_1 =1$ and $g_2 =1$. {}From Fig.~\ref{fig-5}, 
we see that there does not exist the oscillating behavior of $a$, 
and that crossings of the phantom divide can be realized. 

Now, we consider the case that two periods of the elliptic functions are equal 
to infinity, i.e., $m_1=m_2=\infty$, the elliptic functions are 
reduced (degenerate) to the elementary rational functions 
in \eqref{eq:3.7}. 
In this degenerate case, the MG-III model is reduced to the model 
in which the Hubble parameter and the scale factor are given by 
\begin{equation}
H=t\,, 
\quad 
a=a_0e^{0.5t^2}\,. 
\label{eq:4.15}
\end{equation}
Moreover, 
the parametric EoS becomes 
\begin{equation}
\rho = 3t^2\,, 
\quad 
P=-2-3t^2\,. 
\label{eq:4.16}
\end{equation}
It follows from these equations that the EoS is written as 
\begin{equation}
P=-2-\rho\,.
\label{eq:4.17}
\end{equation}
Therefore, the corresponding EoS parameter reads 
\begin{equation}
w =-1-\frac{2}{3t^2}\,. 
\label{eq:4.18}
\end{equation}
Consequently, 
Eq.~\eqref{eq:4.18} informs us that the EoS parameter is always less than 
$-1$ as $w<-1$ and in the limit of $t \to \infty$, $w=-1$. 
Thus, the late-time accelerated expansion of the universe can be realized. 

\section{Models induced by the Jacobian elliptic functions}

In this section, we investigate 
FLRW models (MG-V, MG-VI, $\dots$, 
MG-XVI) induced by the Jacobian elliptic functions 
$\mbox{cn}t \equiv \mbox{cn}(t,m)$, $\mbox{sn}t \equiv \mbox{sn}(t,m)$ 
and $\mbox{dn}t \equiv \mbox{dn}(t,m)$, where $m$ is the parameter of 
the elliptic modulus. 
The Jacobian elliptic functions are doubly periodic functions.  
Hence, 
these functions 
lead to a new class of cosmological models of the cyclic universes with 
a two periodic feature. 
With the same procedure as that in Secs.~III and IV, 
we reconstruct the EoS. 

\subsection{Formulations}

First, we present 
some differential equations for the Jacobian elliptic 
functions~\cite{Book-elliptic-functions}
\begin{eqnarray}
	\mbox{sn}^{\prime }t\Eqn{=}\mbox{cn}t\mbox{dn}t\,,\\
	\mbox{cn}^{\prime}t\Eqn{=}-\mbox{sn}t\mbox{dn}t\,,\\
	\mbox{dn}^{\prime}t\Eqn{=}-m^2\mbox{sn}t\mbox{cn}t\,,\\
%
%
\mbox{sn}^{\prime \prime}t\Eqn{=}-\mbox{sn}t\mbox{dn}^2t-m^2\mbox{sn}t\mbox{cn}^2t=-\mbox{sn}t(\mbox{dn}^2t+m^2\mbox{cn}^2t)\,,\\
\mbox{cn}^{\prime\prime}t\Eqn{=}\mbox{cn}t(m^2\mbox{sn}^2t-\mbox{dn}^2t)\,,\\
\mbox{dn}^{\prime\prime}t\Eqn{=}m^2\mbox{dn}t(\mbox{sn}^2t-\mbox{cn}^2t)\,,
\end{eqnarray}
where $\mbox{cn}^{\prime}t=d\mbox{cn} t/dt$ and so on. 
Hence, we get some known equations for these functions. For example, 
the differential equations for $y = \mbox{sn}t$ 
reads~\cite{Book-elliptic-functions} 
\begin{eqnarray}
&&
\frac{\mathrm{d}^2 y}{\mathrm{d}t^2} + (1+m^2) y - 2 m^2 y^3 = 0\,, \\ 
%
%
&&
\left(\frac{\mathrm{d} y}{\mathrm{d}t}\right)^2 = (1-y^2) (1-m^2 y^2)\,. 
\end{eqnarray}
For $y = \mbox{cn}t$, we have 
the differential equations~\cite{Book-elliptic-functions}  
\begin{eqnarray}
&&
\frac{\mathrm{d}^2 y}{\mathrm{d}t^2} + (1-2m^2) y + 2 m^2 y^3 = 0\,, \\ 
%
%
&&
\left(\frac{\mathrm{d} y}{\mathrm{d}t}\right)^2 = (1-y^2) (1-m^2 + m^2 y^2)\,. 
\end{eqnarray}
%
In addition, 
we present the differential equations for the function 
$y=\mbox{dn}t$~\cite{Book-elliptic-functions} 
\begin{eqnarray}
&&
\frac{\mathrm{d}^2 y}{\mathrm{d}t^2} - (2 - m^2) y + 2 y^3 = 0\,, \\ 
%
%
&&
\left(\frac{\mathrm{d} y}{\mathrm{d}t}\right)^2 = (y^2 - 1) (1 - m^2 - y^2)\,. 
\end{eqnarray}
We note that these functions are doubly periodic generalizations of the trigonometric functions and give them for the particular values of the parameter 
$m$: 
\begin{equation}
	\mbox{sn}(t,0) = \sin(t)\,, \quad 
	\mbox{cn}(t,0) = \cos(t)\,, \quad 
	\mbox{dn}(t,0) = 1\,.
\end{equation}
Also, we present other degenerate values of these functions: 
\begin{equation}
	\mbox{sn}(t,1) = \tan(t)\,, \quad 
	\mbox{cn}(t,1) = \sec(t)\,, \quad 
	\mbox{dn}(t,1) = \sec(t)\,.
\end{equation}
In what follows, we demonstrate several examples of the cyclic cosmological 
models induced by the Jacobian elliptic functions.

\subsection{MG-V model}

As an example, for MG-V model, 
the scale factor is represented as 
\begin{equation}
a(t)=a_0\e^{\mbox{cn}(t)}\,.
\label{eq:5.19}
\end{equation} 
The Hubble parameter is described by
\begin{equation}
H=\mbox{cn}^{\prime}(t)=-\mbox{sn}(t)\mbox{dn}(t)\,.
\end{equation}
With the gravitational field equations,  
the parametric EoS is expressed as  
\begin{eqnarray}
\rho\Eqn{=}3(\mbox{sn}(t)\mbox{dn}(t))^2\,,
\label{eq:5.22} \\
P\Eqn{=}-3\mbox{dn}^2(t)\mbox{sn}^2(t)+2\mbox{cn}(t)\left(\mbox{dn}^2(t)-n\mbox{sn}^2(t)\right)\,. 
\label{eq:5.21} 
\end{eqnarray}
The EoS parameter is given by 
\begin{equation}
w=-1+\frac{2}{3}\mbox{cn}(t)\left(-\frac{n}{\mbox{dn}^2(t)}+\frac{1}{\mbox{sn}^2(t)}\right)\,,
\label{eq:5.23}
\end{equation}
where we have used $n$-elliptic modulus.

\begin{center}
\begin{figure}[t]
\resizebox{!}{6.5cm}{
   \includegraphics{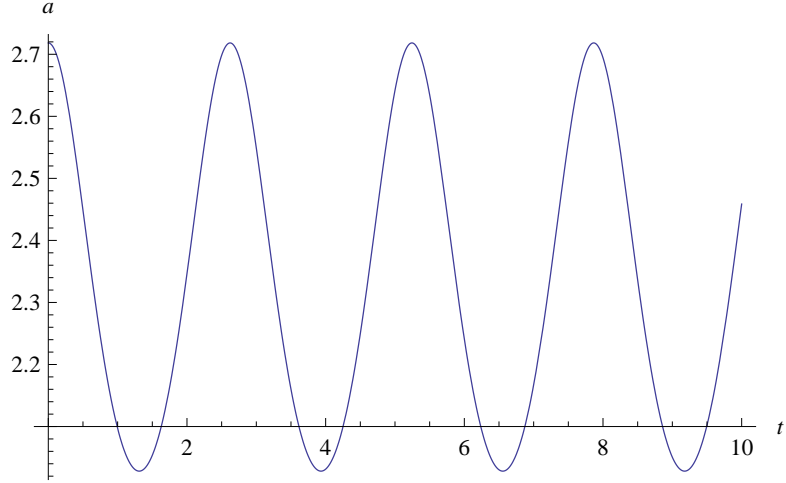}
                  }
\caption{The scale factor $a$ in Eq.~(\ref{eq:5.19}) as a function of $t$ in the MG-V model for the elliptic modulus of $m=2$.}
\label{fig-6}
\end{figure}
\end{center}

\begin{center}
\begin{figure}[t]
\begin{tabular}{ll}
\begin{minipage}{80mm}
\begin{center}
\unitlength=1mm
\resizebox{!}{4.5cm}{
   \includegraphics{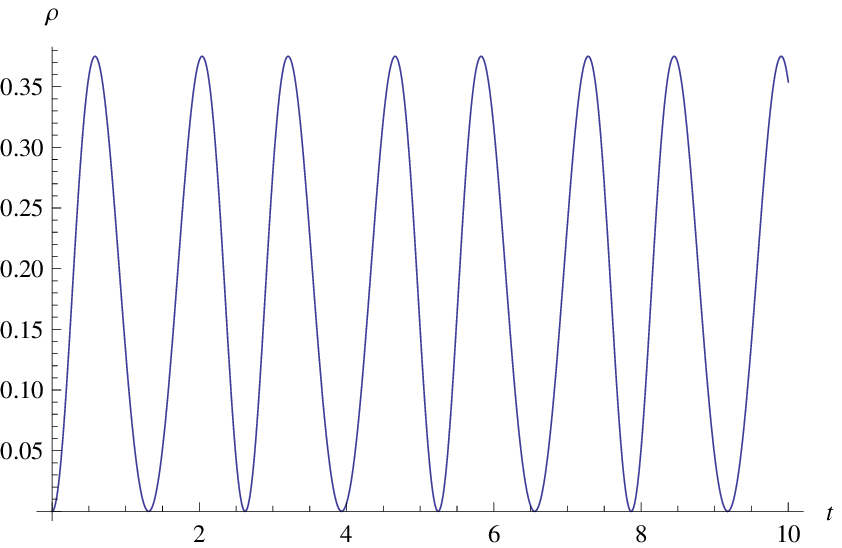}
                  }
\end{center}
\end{minipage}
&
\begin{minipage}{80mm}
\begin{center}
\unitlength=1mm
\resizebox{!}{4.5cm}{
   \includegraphics{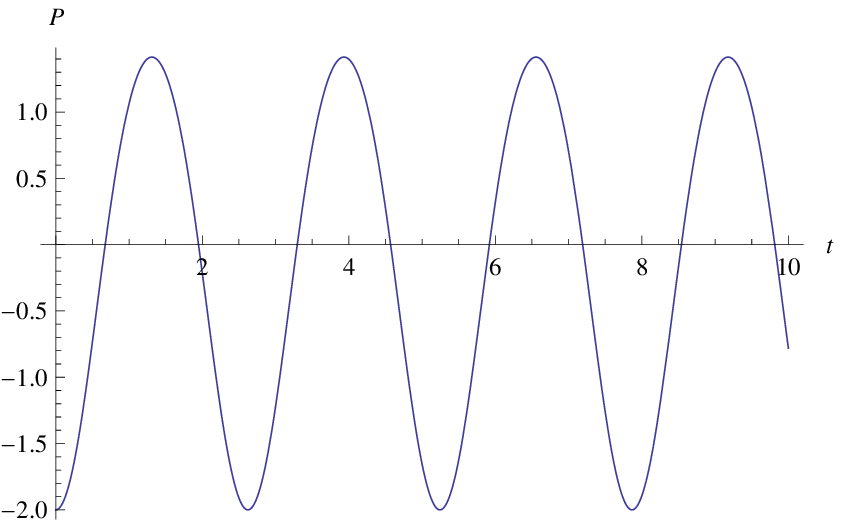}
                  }
\end{center}
\end{minipage}

\end{tabular}
\caption{The energy density $\rho$ in Eq.~(\ref{eq:5.22}) [left panel] 
and pressure $P$ in Eq.~(\ref{eq:5.21}) [right panel] 
as functions of $t$. Legend is the same as Fig.~\ref{fig-6}.
}
\label{fig-7}
\end{figure}
\end{center}

\begin{center}
\begin{figure}[t]
\resizebox{!}{6.5cm}{
   \includegraphics{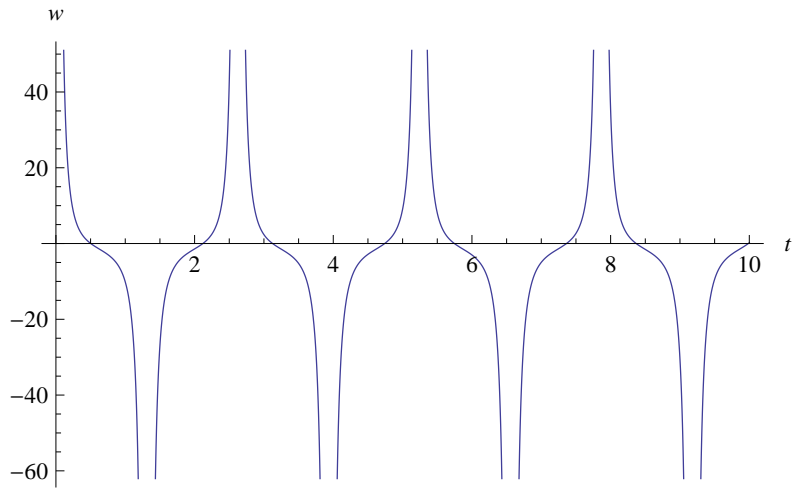}
                  }
\caption{The EoS $w$ in Eq.~(\ref{eq:5.23}) as a function of $t$. Legend is the same as Fig.~\ref{fig-6}.}
\label{fig-8}
\end{figure}
\end{center}

We demonstrate the cosmological evolution of the scale factor 
$a(t)$ in Eq.~(\ref{eq:5.19}) as a function of $t$ in Fig.~\ref{fig-6}. We also show the cosmological evolutions of the energy density $\rho$ in Eq.~(\ref{eq:5.22}) and pressure $P$ in Eq.~(\ref{eq:5.21}) as functions of $t$ in Fig.~\ref{fig-7}. 
Furthermore, in Fig.~\ref{fig-8} we illustrate the cosmological evolution of 
the EoS $w$ in Eq.~(\ref{eq:5.23}) as a function of $t$. 
Here, we have taken the parameter of the elliptic modulus parameter as $m=2$. 
{}From Fig.~\ref{fig-6}, we see the oscillation of $a$. 
Thus, it is considered that by using the Jacobian elliptic functions, a model with realizing the cyclic universe can be reconstructed. 
Moreover, it is clearly seen from Fig.~\ref{fig-8} that crossings of the phantom divide can be realized. 

In Appendix A, 
for other FLRW models (MG-VI, $\dots$, 
MG-XVI), we present the scale factor, the energy density and pressure, and 
the parametric EoS as well as the EoS parameter. 
We have executed the same analysis as that shown in this section. 
In each model, we have examined the cosmological evolutions of the scale factor and the EoS as functions of $t$.

\section{Conclusions}

In the present paper, we have explored 
the cosmological evolutions of the EoS for the 
universe in the homogeneous and isotropic FLRW background. 
With the Weierstrass $\wp(t)$, $\zeta(t)$ and $\sigma(t)$-functions and the Jacobian elliptic functions, we have reconstructed cosmological models which can describe the cyclic universes. 
Furthermore, we have explicitly demonstrated that 
in the MG-I, MG-XI, 
MG-XXXIV, 
MG-XXXVI, MG-XXXVII, MG-XXXVIII models, the universe always stays in the non-phantom (quintessence) phase, 
whereas in the MG-II, MG-III, MG-V, MG-VI, MG-VII, MG-VIII, MG-IX, MG-X, MG-XXXV models, the crossing of the phantom divide can be realized. 

It is known that 
there exist two approaches to produce the cyclic universes. 
One is to introduce non-canonical scalar field, which makes the vacuum 
unstable in the quantum field theory. Another is to extend the gravitational theory, such as gravity with the higher order derivative term and $f(R)$ gravity. 
It is also important to indicate that our MG-i (i=I, II, $\cdots$) models are 2-periodic or 1-periodic generalizations of the usual non-periodic FLRW models. 

We note that 
the issue of removing singularities is a fundamental question. 
In other words, this is what a new physics comes out, when in the cosmological models the energy/curvature scale is so high that render the hypothesis of general relativity should be rendered inapplicable. 
Although this is a question far from a satisfactory solution, a complete picture of these cosmologies certainly will need to take this issue seriously into account. 
Furthermore, 
there exists models, in particular, based on higher-curvature theories, 
with possessing bouncing solutions, i.e., without curvature singularities, 
that modify general relativity only if the universe becomes close to 
the predicted singularity, e.g., at the Planck scale. 
On the one hand, these theories would pass the solar test equally well as 
general relativity. 
Moreover, in principle, 
the perspective of cyclic cosmologies could be realized without 
supposing the existence of singularities as an unavoidable conclusion of the 
theory. 
This issue should be considered in more detailed in the future works. 

In Ref.~\cite{Bamba:2012cp}, it has been verified that 
all the dark energy cosmologies can be realized by different fluids 
of the universe with a general form of the EoS, 
and that 
the evolutions of all the dark energy universes at the present time can 
be similar to that of the $\Lambda$CDM model, in which the universe consists 
of the cosmological constant $\Lambda$ and cold dark matter (CDM). 
Since the $\Lambda$CDM model is compatible with the various cosmological observations~\cite{WMAP, SN1, LSS, Eisenstein:2005su, Jain:2003tba}, 
the models of the dark energy fluids are also consistent with the current 
observations. 
Indeed, it has been illustrated that 
different dark energy models are equivalent by examining various 
scalar field theories such as single and multiple scalar fields theories 
as well as tachyon scalar theory and holographic dark energy. 
In these theories, the current accelerated expansion of the universe 
with the quintessence or phantom phases can be realized. 
Also, the equivalence of these theories 
to the corresponding fluid descriptions have been shown. 
According to these consequences, 
we can understand that for 
the MG-I, MG-XI, 
MG-XXXIV, MG-XXXVI, MG-XXXVII, MG-XXXVIII models, 
the non-phantom (quintessence) phase occurs 
and hence these fluid models correspond to the quintessence models. 
On the other hand, for 
the MG-II, MG-III, MG-V, MG-VI, MG-VII, MG-VIII, MG-IX, MG-X, MG-XXXV models, 
since the crossing of the phantom divide can happen, 
these fluid models can be regarded to be equivalent to 
the two scalar fields models, for example, 
the quintom models~\cite{Cai:2009zp, Quintom} with 
both the canonical and non-canonical scalar fields. 

The new cosmological consequence obtained in this work is that 
by combining the reconstruction method of the EoS for the universe 
with mathematical special functions 
such as the Weierstrass and Jacobian elliptic functions, 
new descriptions of the cyclic universes can be acquired. 
It is considered that this can be worthy of a clue of one of the generalized 
describing manner of the expansion history of the universe. 

In addition, it is remarkable to mention that if we apply the investigations 
in this work to the spatially anisotropic space-time, the EoS for the universe 
has the asymmetry during the oscillatory process of the expansion and contraction, so that the cosmological hysteresis can be realized~\cite{Sahni:2012er}. 

Finally, it is also important to stress that 
concerning the issue on potential deviation of a gravitational theory from general relativity, the detection of gravitational waves could be a definitive tool 
for discrimination between a gravitational theory and general relativity, 
as shown in Ref.~\cite{Corda:2009re}.

\section*{Acknowledgments}

K.B. is sincerely grateful to all of the members of Eurasian International Center for Theoretical Physics for their very kind and warm hospitality, where this work has been completed.

\appendix
\section{Other models} 

In this appendix, we describe 
other FLRW models (MG-VI, $\dots$, 
MG-XVI). 
For each model, 
by executing the numerical calculations for the parameter of the elliptic modulus of $m=2$, we have analyzed the cosmological evolutions of the scale factor 
and the EoS.

\subsection{MG-VI model}

The scale factor, the energy density and pressure, and 
the parametric EoS, and the EoS parameter are given by 
\begin{eqnarray}
a(t) \Eqn{=} a_0\e^{\mbox{sn}(t)}\,, 
\label{eq:V-a} \\
H \Eqn{=} \mbox{sn}^{\prime}(t)=\mbox{dn}(t)\mbox{cn}(t)\,, \\
\rho\Eqn{=}3[\mbox{cn}(t)\mbox{dn}(t)]^2\,, \\ 
P\Eqn{=}-3\mbox{dn}^2(t)\mbox{cn}^2(t)-2\mbox{sn}(t)\left(-\mbox{dn}^2(t)-n\mbox{cn}^2(t)\right)\,,\\
w \Eqn{=} -1+\frac{2}{3}\mbox{sn}(t)\left(\frac{n}{\mbox{dn}^2(t)}+\frac{1}{\mbox{cn}^2(t)}\right)\,.
\label{eq:V-w}
\end{eqnarray}
We have confirmed the oscillation of $a$ and found 
that crossings of the phantom divide can be realized.

\subsection{MG-VII model}
 
The scale factor, the energy density and pressure, and 
the parametric EoS, and the EoS parameter are given by 
\begin{eqnarray}
a(t) \Eqn{=} a_0\e^{\mbox{dn}(t)}\,, 
\label{eq:VI-a} \\
H \Eqn{=} \mbox{dn}^{\prime}t\,, \\
\rho\Eqn{=}3[n\mbox{cn}(t)\mbox{sn}(t)]^2\,, \\
P\Eqn{=}-3n^2\mbox{sn}^2(t)\mbox{cn}^2(t)+2n\mbox{dn}(t)\left(\mbox{cn}^2(t)-\mbox{sn}^2(t)\right)\,,\\
w \Eqn{=} -1-\frac{2\mbox{dn}(t)}{3n}\left(\frac{1}{\mbox{cn}^2(t)}-\frac{1}{\mbox{sn}^2(t)}\right)\,.
\label{eq:VI-w}
\end{eqnarray}
We have confirmed the oscillation of $a$ and found 
that crossings of the phantom divide can be realized.

\subsection{MG-VIII model}

The scale factor, the energy density and pressure, and 
the parametric EoS, and the EoS parameter are given by 
\begin{eqnarray}
a(t)\Eqn{=}a_0\exp \left( \frac{\arccos[\mbox{dn}(t)]\mbox{sn}(t)}{\sqrt[2]{1-\mbox{dn}^2(t)}} \right)\,, 
\label{eq:VII-a} \\
H\Eqn{=}\mbox{cn}t\,, \\ 
\rho\Eqn{=}3\mbox{cn}^2(t)\,, \\ 
P\Eqn{=}-3\mbox{cn}^2(t)+2\mbox{dn}(t)\mbox{sn}(t)\,,\\
w\Eqn{=}-1+\frac{2\mbox{dn}(t)\mbox{sn}(t)}{3\mbox{cn}^2(t)}\,.
\label{eq:VII-w}
\end{eqnarray}
We have confirmed that there is no oscillating behavior of $a$ 
and found that crossings of the phantom divide can be realized.

\subsection{MG-IX model}

The scale factor, the energy density and pressure, and 
the parametric EoS, and the EoS parameter are given by 
\begin{eqnarray}
a(t) \Eqn{=} a_0(-\sqrt{n}\mbox{cn}(t)+\mbox{dn}(t))^{1/\sqrt{n}}\,.
\label{eq:VIII-a} \\ 
H \Eqn{=} \mbox{sn}(t)\,, \\ 
\rho\Eqn{=}3\mbox{sn}^2(t)\,, \\ 
P\Eqn{=}-3\mbox{sn}^2(t)-2\mbox{dn}(t)\mbox{cn}(t)\,,\\ 
w \Eqn{=} -1-\frac{2\mbox{dn}(t)\mbox{cn}(t)}{3\mbox{sn}^2(t)}\,.
\label{eq:VIII-w}
\end{eqnarray}
We have seen that crossings of the phantom divide can be realized.

\subsection{MG-X model}

The scale factor, the energy density and pressure, and 
the parametric EoS, and the EoS parameter are given by 
\begin{eqnarray}
a(t) \Eqn{=} a_0\e^{am(t)}\,,
\label{eq:IX-a} \\ 
H \Eqn{=} \mbox{dn}t\,, \\ 
\rho\Eqn{=}3\mbox{dn}^2(t)\,, \\ 
P\Eqn{=}-3\mbox{dn}^2(t)+2n\mbox{sn}(t)\mbox{cn}(t)\,, \\
w \Eqn{=} -1+\frac{2n\mbox{sn}(t)\mbox{cn}(t)}{3\mbox{dn}^2(t)}\,.
\label{eq:IX-w}
\end{eqnarray}
We have confirmed the oscillation of $a$ 
and found that crossings of the phantom divide can be realized.

\subsection{MG-XXXIII model}

The scale factor, the energy density and pressure, and 
the parametric EoS, and the EoS parameter are given by 
\begin{eqnarray}
a(t) \Eqn{=} \mbox{cn}^{\prime}(t)=-\mbox{dn}(t)\mbox{sn}(t)\,.
\label{eq:X-a} \\ 
H \Eqn{=} \mbox{cn}t \left(\frac{\mbox{dn}(t)}{\mbox{sn}(t)}-n\frac{\mbox{sn}(t)}{\mbox{dn}(t)}\right)\,, \\ 
\rho\Eqn{=}3\left(\mbox{cn}t\left(\frac{\mbox{dn}(t)}{\mbox{sn}(t)}-n\frac{\mbox{sn}(t)}{\mbox{dn}(t)}\right)\right)^2\,, \\
P\Eqn{=}-3\left[\mbox{cn}(t)\left(\frac{\mbox{dn}(t)}{\mbox{sn}(t)}-n\frac{\mbox{sn}(t)}{\mbox{dn}(t)}\right)\right]^2
\nonumber \\
&&
{}-2\left[\frac{-\mbox{dn}^4(t)\mbox{sn}^2(t)+n\mbox{dn}^2(t)\mbox{sn}^4(t)
-\mbox{cn}^2(t)\left(\mbox{dn}^2(t)+n\mbox{sn}^2(t)\right)^2}{\mbox{dn}^2(t)
\mbox{sn}^2(t)}\right]\,, \\ 
w \Eqn{=} -1-\frac{2\left[\frac{-\mbox{dn}^4(t)\mbox{sn}^2(t)+n\mbox{dn}^2(t)\mbox{sn}^4(t)-\mbox{cn}^2(t)\left(\mbox{dn}^2(t)+n\mbox{sn}^2(t)\right)^2}{\mbox{dn}^2(t)\mbox{sn}^2(t)}\right]}{3\left[\mbox{cn}(t)\left(\frac{\mbox{dn}(t)}{\mbox{sn}(t)}-n\frac{\mbox{sn}(t)}{\mbox{dn}(t)}\right)\right]^2}\,.
\label{eq:X-w}
\end{eqnarray}
We have confirmed the oscillation of $a$ 
and found that the universe always stays in the non-phantom (quintessence) 
phase ($w > -1$).

\subsection{MG-XXXIV model}

The scale factor, the energy density and pressure, and 
the parametric EoS, and the EoS parameter are given by 
\begin{eqnarray}
a(t) \Eqn{=} \mbox{sn}^{\prime}(t)=\mbox{cn}(t)\mbox{dn}(t)\,, 
\label{eq:XI-a} \\ 
H \Eqn{=} -\frac{\mbox{sn}(t)\left(n\mbox{cn}^2(t)+\mbox{dn}^2(t)\right)}{\mbox{cn}(t)\mbox{dn}(t)}\,, \\  
\rho\Eqn{=}3\frac{\mbox{sn}^2(t)\left(n\mbox{cn}^2(t)+\mbox{dn}^2(t)\right)^2}{\mbox{cn}^2(t)\mbox{dn}^2(t)}\,, \\ 
P\Eqn{=}-3\frac{\mbox{sn}^2(t)\left(n\mbox{cn}^2(t)+\mbox{dn}^2(t)\right)^2}{\mbox{cn}^2(t)\mbox{dn}^2(t)}
\nonumber \\
&&
{}-2\left[-\mbox{dn}^2(t)+\mbox{sn}^2(t)\left(2n-\frac{\mbox{dn}^2(t)}{\mbox{cn}^2(t)}\right)+n\mbox{cn}^2(t)\left(-1-n\frac{\mbox{sn}^2(t)}{\mbox{dn}^2(t)}\right)\right]\,,\\ 
w\Eqn{=}-1-\frac{2}{3}\frac{\left[-\mbox{dn}^2(t)+\mbox{sn}^2(t)\left(2n-\frac{\mbox{dn}^2(t)}{\mbox{cn}^2(t)}\right)+n\mbox{cn}^2(t)\left(-1-n\frac{\mbox{sn}^2(t)}{\mbox{dn}^2(t)}\right)\right]}{\mbox{sn}^2(t)\frac{n^2(\mbox{cn}^2(t)+\mbox{dn}^2(t))^2}{\mbox{dn}^2(t)\mbox{cn}^2(t)}}\,.
\label{eq:XI-w}
\end{eqnarray}
We have confirmed the oscillation of $a$ 
and found that 
the universe always stays in the non-phantom (quintessence) 
phase ($w > -1$).

\subsection{MG-XXXV model}

The scale factor, the energy density and pressure, and 
the parametric EoS, and the EoS parameter are given by 
\begin{eqnarray}
a(t) \Eqn{=} -n\mbox{cn}(t)\mbox{sn}(t)\,, 
\label{eq:XII-a} \\  
H \Eqn{=} \frac{\mbox{cn}(t)\mbox{dn}(t)}{\mbox{sn}(t)}-\frac{\mbox{sn}(t)\mbox{dn}(t)}{\mbox{cn}(t)}\,, \\ 
\rho\Eqn{=}3\left(\frac{\mbox{cn}(t)\mbox{dn}(t)}{\mbox{sn}(t)}-\frac{\mbox{sn}(t)\mbox{dn}(t)}{\mbox{cn}(t)}\right)^2\,, \\ 
P\Eqn{=}-3\left(\frac{\mbox{cn}(t)\mbox{dn}(t)}{\mbox{sn}(t)}-\frac{\mbox{sn}(t)\mbox{dn}(t)}{\mbox{cn}(t)}\right)^2
\nonumber \\
&&
{}-2\left[-2\mbox{dn}^2(t)+\mbox{cn}^2(t)\left(-n-\frac{\mbox{dn}^2(t)}{\mbox{sn}^2(t)}\right)+\mbox{sn}^2(t)\left(n-\frac{\mbox{dn}^2(t)}{\mbox{cn}^2(t)}\right)\right]\,,\\ 
w \Eqn{=} -1-\frac{2\left[-2\mbox{dn}^2(t)+\mbox{cn}^2(t)\left(-n-\frac{\mbox{dn}^2(t)}{\mbox{sn}^2(t)}\right)+\mbox{sn}^2(t)\left(n-\frac{\mbox{dn}^2(t)}{\mbox{cn}^2(t)}\right)\right]}{3\left(\frac{\mbox{cn}(t)\mbox{dn}(t)}{\mbox{sn}(t)}-\frac{\mbox{sn}(t)\mbox{dn}(t)}{\mbox{cn}(t)}\right)^2}\,.
\label{eq:XII-w}
\end{eqnarray}
We have confirmed the oscillation of $a$ and 
found that the crossings of the phantom divide can be realized.

\subsection{MG-XXXVI model}

The scale factor, the energy density and pressure, and 
the parametric EoS, and the EoS parameter are given by 
\begin{eqnarray}
a(t) \Eqn{=} \mbox{cn}(t)\,, 
\label{eq:XIII-a} \\ 
H \Eqn{=} -\frac{\mbox{sn}(t)\mbox{dn}(t)}{\mbox{cn}(t)}\,, \\ 
\rho \Eqn{=} 3\left(\frac{\mbox{sn}(t)\mbox{dn}(t)}{\mbox{cn}(t)}\right)^2\,, \\ 
P\Eqn{=}-3\left(\frac{\mbox{sn}(t)\mbox{dn}(t)}{\mbox{cn}(t)}\right)^2-2\left[n\mbox{sn}^2(t)+\mbox{dn}^2(t)\left(-1-\frac{\mbox{sn}^2(t)}{\mbox{cn}^2(t)}\right)\right]\,, \\
w \Eqn{=} -1-\frac{2\left[n\mbox{sn}^2(t)+\mbox{dn}^2(t)\left(-1-\frac{\mbox{sn}^2(t)}{\mbox{cn}^2(t)}\right){\mbox{cn}^2(t)}\right]}{3\left(\frac{\mbox{sn}(t)\mbox{dn}(t)}{\mbox{cn}(t)}\right)^2}\,. 
\label{eq:XIII-w}
\end{eqnarray}
We have confirmed the oscillation of $a$ and found that 
the universe always stays in the non-phantom (quintessence) 
phase ($w > -1$).

\subsection{MG-XXXVII model}

The scale factor, the energy density and pressure, and 
the parametric EoS, and the EoS parameter are given by 
\begin{eqnarray}
a(t) \Eqn{=} \mbox{sn}(t)\,, 
\label{eq:XIV-a} \\ 
H \Eqn{=} \frac{\mbox{cn}(t)\mbox{dn}(t)}{\mbox{sn}(t)}\,, \\ 
\rho\Eqn{=}3\left(\frac{\mbox{cn}(t)\mbox{dn}(t)}{\mbox{sn}(t)}\right)^2\,, \\ 
P\Eqn{=}-3\left(\frac{\mbox{cn}(t)\mbox{dn}(t)}{\mbox{sn}(t)}\right)^2-2\left[-\mbox{dn}^2(t)+\mbox{cn}^2(t)\left(-n-\frac{\mbox{dn}^2(t)}{\mbox{sn}^2(t)}\right)\right]\,, \\ 
w \Eqn{=} -1-\frac{2\left[-\mbox{dn}^2(t)+\mbox{cn}^2(t)\left(-n-\frac{\mbox{dn}^2(t)}{\mbox{sn}^2(t)}\right)\right]}{3(\frac{\mbox{cn}(t)\mbox{dn}(t)}{\mbox{sn}(t)})^2}\,.
\label{eq:XIV-w}
\end{eqnarray}
We have confirmed the oscillation of $a$ 
and found that the universe always stays in the non-phantom (quintessence) 
phase ($w > -1$).

\subsection{MG-XXXVIII model}

The scale factor, the energy density and pressure, and 
the parametric EoS, and the EoS parameter are given by 
\begin{eqnarray}
a(t) \Eqn{=} \mbox{dn}(t)\,, 
\label{eq:XV-a} \\ 
H \Eqn{=} -n\frac{\mbox{cn}(t)\mbox{sn}(t)}{\mbox{dn}(t)}\,, \\ 
\rho\Eqn{=}3\left(n\frac{\mbox{cn}(t)\mbox{sn}(t)}{\mbox{dn}(t)}\right)^2\,, \\
P\Eqn{=}-3\left(n\frac{\mbox{cn}(t)\mbox{sn}(t)}{\mbox{dn}(t)}\right)^2-2n\left[\mbox{sn}^2(t)+\mbox{cn}^2(t)\left(-1-n\frac{\mbox{sn}^2(t)}{\mbox{dn}^2(t)}\right)\right]\,, \\ 
w \Eqn{=} -1-\frac{2n\left[ \mbox{sn}^2(t)+\mbox{cn}^2(t)\left(-1-n\frac{\mbox{sn}^2(t)}{\mbox{dn}^2(t)}\right)\right]}{3\left(n\frac{\mbox{cn}(t)\mbox{sn}(t)}{\mbox{dn}(t)}\right)^2}\,. 
\label{eq:XV-w} 
\end{eqnarray}
We have confirmed the oscillation of $a$ 
and found that 
the universe always stays in the non-phantom (quintessence) 
phase ($w > -1$).


\end{document}